\documentclass[aps,prd,twocolumn,showpacs,groupedaddress,floatfix,nofootinbib]{revtex4}
\usepackage{amsmath}
\usepackage{graphicx}
\usepackage{color}
\newcommand{\beq}{\begin{eqnarray}}
\newcommand{\eeq}{\end{eqnarray}}

\begin{document}

\title{{Non perturbative regularization of one--loop integrals at finite temperature}}
\author{Paolo Amore}
\email{paolo@ucol.mx}
\affiliation{Facultad de Ciencias, Universidad de Colima,\\
Bernal D\'{i}az del Castillo 340, Colima, Colima,\\
Mexico.}

\begin{abstract}
A method devised by the author is used to calculate analytical expressions for one--loop integrals
at finite temperature. A non-perturbative regularization of the integrals is performed, yielding expressions
of non-polynomial nature. A comparison with previously published results is 
presented and the advantages of the present technique are discussed. 
\end{abstract}


\maketitle

\section{Introduction}

In this paper we present a new technique which allows to evaluate {\sl non-perturbatively} one--loop 
integrals occurring in finite temperature problems in quantum field theory. Such integrals typically involve 
series, due to summation over the the Matsubara frequencies, $\omega_n = 2 \pi n T$, with 
$n =0, \pm 1, \dots$~\cite{FW71}, which cannot be done analytically.

The method that we propose here is derived from the powerful ideas of the Linear Delta Expansion (LDE) and 
of Variational Perturbation Theory (VPT) (see \cite{LDEVPT} and references therein); 
we briefly sketch how such methods work.

First one interpolates a non-perturbative lagrangian (hamiltonian) ${\cal L}$ with a solvable one 
${\cal L}_0$, which depends on one or more arbitrary parameters; the interpolated lagragian is then 
split into a ``leading'' term, chosen to be ${\cal L}_0$ itself and into a ``perturbation'', 
${\cal L}-{\cal L}_0$. Although the ``perturbative'' term will not be a-priori small, a perturbative 
expansion is carried out and physical quantities are calculated. To finite order in perturbation 
theory one thus obtains expression which  contain an artificial dependence upon the arbitrary 
parameters: however if the expansion were to be carried out to all orders such dependence would have to cancel 
out provided that the perturbative series converges. In order to minimize such dependence one applies the
Principle of Minimal Sensitivity (PMS)~\cite{Ste81}, which selects the value of the arbitrary parameter for
which the physical quantity is less sensitive to changes in the parameter itself. The parameter determined 
in this way is normally such to make ${\cal L}-{\cal L}_0$ a perturbation and depends on the ``natural'' parameters
present in ${\cal L}$. For this reason, the expansion obtained by using this method does not provide, 
to a finite order, a polynomial in the parameters of the model, as  would be the case for a genuinely
perturbative technique. 

The ideas behind the LDE and the VPT are really very general and have been successful in dealing with a variety of 
problems of quite different nature, ranging from quantum field theory to classical and quantum mechanics. 
In this paper, in particular, we pursue the application of some results which were recently obtained by the author~\cite{Am_zeta04}: 
in that paper the author developed a method to accelerate the convergence of 
certain class of mathematical series (including the Riemann and Hurwitz zeta functions) and proved that the new series 
obtained in such a way converge exponentially to the correct results. As we will see in this paper, some of the series 
treated in \cite{Am_zeta04} are relevant for the evaluation of one--loop integrals at finite temperature and will be used
to obtain arbitrarily precise analytical approximations to such integrals which are not polynomials in the inverse temperature.

The paper is organized as follows: in section \ref{method} we review some of the results of \cite{Am_zeta04} and explain the method;
in section \ref{appl} we apply the results of section  \ref{method} to the calculation of one--loop integrals at finite temperature 
and compare them with the results in the literature~\cite{DJ74,LV97,AE93,HW82}; finally in section \ref{concl} we draw our conclusions
and discuss further possible applications of this method.

\section{The method}
\label{method}

In this section we describe the method of \cite{Am_zeta04}, which allows one to accelerate the convergence of certain 
series. One of the applications which was discussed in that paper is to the zeta-like function defined as 
\beq
\zeta(u,s,\xi) \equiv \sum_{n=0}^\infty \frac{1}{(n^u+\xi)^s}  \ ,
\label{ap_1_1}
\eeq
where $s$, $u$ and $\xi$ are real parameters. The special cases obtained by taking $(u,\xi)=(1,1)$ or 
$u=1$ correspond to the Riemann and Hurwitz zeta functions respectively. 
Further examples of application of this method are contained in \cite{Am_zeta04} and the interested 
reader should refer to that paper for further reading.

We write
\begin{eqnarray}
\zeta(u,s,\xi) &=& \frac{1}{\xi^s} + \sum_{n=1}^\infty \frac{1}{\left(n^u + \xi \right)^s}   
\label{ap_1_2}
\end{eqnarray}
and by simple algebra we convert it to the form
\begin{eqnarray}
\zeta(u,s,\xi)  &=& \frac{1}{\xi^s} + \sum_{n=1}^\infty \frac{1}{n^{s u}} \  \frac{1}{\left(1+\lambda^2\right)^s} 
\frac{1}{\left(1+ \Delta(n)\right)^s} 
\label{ap_1_3}
\end{eqnarray}
having introduced the definition
\begin{eqnarray}
\Delta(n) &\equiv& \frac{\xi/n^u-\lambda^2}{1+\lambda^2} \ .
\label{ap_1_4}
\end{eqnarray}

We stress that eq.~(\ref{ap_1_3}) is still an identity and that no approximation has been made so far. 
$\lambda$ here is an arbitrary parameter which was introduced ``ad hoc'': this procedure is typical of 
the Linear Delta Expansion (LDE) and of Variational Perturbation Technique (VPT)~\cite{LDEVPT}.  

Clearly, when the condition $|\Delta(n)| < 1$ is met, one can expand  eq.~(\ref{ap_1_3}) in powers of $\Delta$;
such a condition requires that  $\left|\frac{\xi-\lambda^2}{1+\lambda^2}\right| < 1$, i.e. that $\lambda^2 > \frac{\xi-1}{2}$.

We can therefore write
\begin{eqnarray}
\frac{1}{(1+\Delta(n))^s} = \sum_{k=0}^\infty \frac{\Gamma(k+s)}{\Gamma(s) \ k!} \ (-\Delta(n))^k
\label{ap_1_5}
\end{eqnarray}
and convert the series  (\ref{ap_1_3}) to the equivalent series:
\begin{widetext}
\begin{eqnarray}
\zeta(u,s,\xi)  &=& \frac{1}{\xi^s} + \sum_{n=1}^\infty \sum_{k=0}^\infty \frac{\Gamma(k+s)}{\Gamma(s) \ k!} \ 
\sum_{j=0}^k \ \left( \begin{array}{c}
k \\
j \\
\end{array} \right) \frac{\lambda^{2 (k-j)}}{(1+\lambda^2)^{s+k}} \frac{(-\xi)^j}{n^{u (s+j)}}  \ .
\label{ap_1_6}
\end{eqnarray}
\end{widetext}

By interchanging the sums and performing the sum over $n$ we finally obtain the expression
\begin{widetext}\begin{eqnarray}
\zeta(u,s,\xi)  &=& \frac{1}{\xi^s} + 
\sum_{k=0}^\infty \frac{\Gamma(k+s)}{\Gamma(s) } \ \sum_{j=0}^k \ \frac{(-\xi)^j}{j! (k-j)!} 
\frac{\lambda^{2 (k-j)}}{(1+\lambda^2)^{s+k}}  \zeta(u (s+j)) \ .
\label{ap_1_7}
\end{eqnarray}
\end{widetext}

We stress that eq.~(\ref{ap_1_7}) is still {\sl exact} for $\lambda^2 > \frac{\xi-1}{2}$ and that it 
converges exponentially. Actually this equation describes an entire family of series, all converging to the 
same function, although with different rates of convergence. When the sum in eq.~(\ref{ap_1_7}) is truncated
to a finite value, a fictitious dependence on the arbitrary parameter $\lambda$ is generated: in order to minimize such dependence
and to obtain a series with an optimal rate of convergence one can apply the Principle of Minimal Sensitivity (PMS)~\cite{Ste81},
corresponding to selecting the value of $\lambda$ for which the derivative of the partial sum with respect to $\lambda$ vanishes. 

To leading order one obtains the result
\begin{eqnarray}
\lambda_{PMS}^{(1)} &=& \sqrt{\xi} \ \sqrt{\frac{\zeta( u (1+ s))}{ \zeta(s u)}} ,
\label{ap_1_8} 
\end{eqnarray}
which can then be used inside eq.~(\ref{ap_1_7}). The precision of these formulas has been investigated in 
\cite{Am_zeta04}, showing that extremely precise results are in general obtained already 
working to low orders.

\section{Applications: one--loop integrals at finite temperature}
\label{appl}

We will follow the notation set in \cite{LV97} and write the expression for a general one--loop integral 
at finite temperature in the form
\beq
J(m,a,b) &=& T \mu^{2 \epsilon}  \sum_{\begin{array}{c}n=-\infty\\ n\neq 0\\ \end{array}}^\infty
\int \frac{d^Dk}{(2 \pi)^D} \ \frac{(k^2)^a}{\left[k^2+\omega_n^2+m^2\right]^b}  \nonumber 
\eeq
$\mu$ being the scale brought in by dimensional regularization, $a$ and $b$ being integers ($a \geq 0$ and $b>0$). 
Following \cite{LV97} we also define
\beq
K^2 \equiv \left( \frac{k}{2 \pi T}\right)^2 \  , \  \ 
M^2 \equiv \left( \frac{m}{2 \pi T}\right)^2 \   , \ \
\Omega^2 \equiv \left( \frac{\mu}{2 \pi T}\right)^2 
\nonumber
\eeq
and obtain
\beq
J(M,a,b) &=& T \ \left(2\pi T\right)^{3+2 a -2 b} \ 2 \Omega^{2 \epsilon} \nonumber \\
&\cdot& \frac{\pi^{D/2}}{(2\pi)^D} \ 
\frac{\Gamma(D/2+a)}{\Gamma(D/2)} \ \frac{\Gamma(l)}{\Gamma(b)} \ S(M,l)
\label{eq_3}
\eeq
where 
\beq
S(M,l) &=& \sum_{n=1}^\infty \ \frac{1}{(n^2+M^2)^l}
\label{eq_4}
\eeq
and $l = b-a-D/2$ and $D= 3 - 2\ \epsilon$. Depending upon the value of $l$, the series of eq.~(\ref{eq_4}) could
be divergent and therefore need regularization. 

Eq.~(\ref{eq_4}) is clearly of the form considered in eq.~(\ref{ap_1_1}) and one can write 
\beq
S(M,l) &=& - \frac{1}{M^{2l}} + \zeta(2,l,M^2) \ ,
\eeq
where
\begin{eqnarray}
\zeta(2,l,M^2)  &=& \frac{1}{M^{2l}} \nonumber \\
&+&  \sum_{k=0}^\infty \frac{\Gamma(k+l)}{\Gamma(l) } \  \Psi_{k}(\lambda,2,l,M^2) 
\label{eqq}
\end{eqnarray}
and 
\beq
\Psi_{k}(\lambda,2,l,M^2) &\equiv& \frac{1}{(1+\lambda^2)^{s+k}} \nonumber \\
&\cdot& \sum_{j=0}^k \ \frac{(-M^2)^j}{j! (k-j)!} 
\lambda^{2 (k-j)} \zeta(2 (l+j)) 
\nonumber \ .
\end{eqnarray}

Once more we stress that although the series all converge to the same result independently of $\lambda$ (provided that 
$\lambda > \frac{\xi-1}{2}$), the partial sums, obtained by truncating the series to a finite order will necessarily 
display a dependence on the parameter. Such dependence, which is an artifact of working to a finite order, will also 
make  the rate of convergence of the different elements of the family $\lambda$-dependent. 
Since it is desirable to obtain the most precise results with the least effort, one will select the optimally convergent 
series by fixing $\lambda$ through the ``principle of minimal sensitivity'' (PMS)~\cite{Ste81}.
The solutions obtained by applying this simple criterion  display in general the highest convergence rate and, once plugged
back in the original series, provide a {\sl non-polynomial} expression in the natural parameters. 

In the present case, the ``natural'' parameter in eq.~(\ref{eq_4}) is $M^2$ and to first--order the PMS
yields
\begin{eqnarray}
\lambda_{PMS}^{(1)} &=& \sqrt{M^2} \ 
 \sqrt{\frac{\zeta( 2 (1+ l))}{ \zeta(2 l)}} ,
\label{s3_7}
\end{eqnarray}

On the other hand, if the value $\lambda = 0$ is chosen, then one obtains the series
\beq
\zeta(2,l,M^2)  &=& \frac{1}{M^{2l}} \nonumber \\
&+&  \sum_{k=0}^\infty \frac{\Gamma(k+l)}{\Gamma(l) } \ 
\frac{(-M^2)^k}{k!}  \ \zeta(2 (l+k )) \ ,
\label{l0}
\end{eqnarray}
which, to a finite order yields a polynomial in $M$. Notice that such series corresponds to the  
expansion used in \cite{LV97}. 

\begin{figure}
\includegraphics[width=8cm]{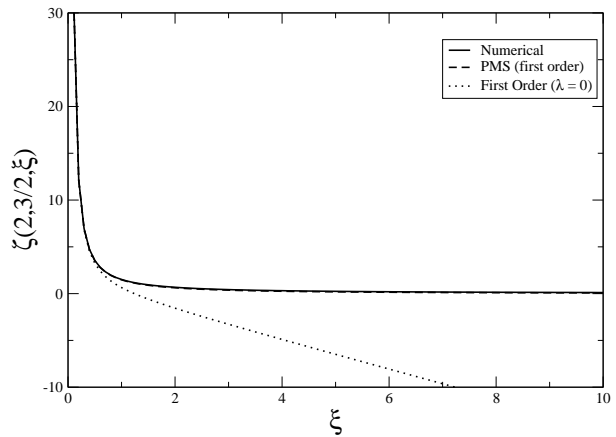}
\caption{$\zeta(2,3/2,\xi)$ calculated numerically (solid line), with the PMS to first order (dashed line) 
and using $\lambda=0$ (dotted line).
 \label{Fig_1}}
\end{figure}

In fig.~\ref{Fig_1} we have compared $\zeta(2,3/2,\xi)$ calculated numerically with the first--order approximations 
obtained by using eq.~(\ref{eqq}) with $\lambda = \lambda_{PMS}^{(1)}$ and $\lambda=0$, which are 
\beq
\zeta_{PMS}^{(1)}(2,3/2,\xi) &=& \frac{1}{\xi^{3/2 }} + \frac{\zeta(3)^{5/2}}{{\left( \zeta(3) + \xi\,\zeta(5) \right) }^{3/2}} \\
\zeta_{\lambda=0}^{(1)}(2,3/2,\xi) &=&  \frac{1}{\xi^{3/2 }} + \zeta(3) - \frac{3}{2} \,\xi\,\zeta(5) \ ,
\eeq
where the second and third term in the second equation can be obtained by Taylor expanding the first equation.

In table \ref{tab:table1} we have calculated $\zeta(2,3/2,1)$ using eq.~(\ref{eqq}) to order $N$ with $\lambda_{PMS}^{(1)}$ (second column) and 
with $\lambda=0$ (third column). Our formula, using the optimal parameter obtained to first order, converges exponentially to the exact 
value, whereas the formula corresponding to $\lambda=0$ is actually useless (that formula is indeed  limited to $\xi \ll 1$).
Notice that $\zeta(2,3/2,1)$ calculated with the first $10^5$ terms in eq.~(\ref{ap_1_1}) gives the result 
$\underline{1.5124349215}0$; the same precision is reached with our improved series with only $27$ terms. We believe that this result
by itself is sufficient to illustrate the strength of our method.

\begin{table*}
\caption{\label{tab:table1} $\zeta(2,3/2,1)$ calculated using eq.~(\ref{eqq}) to order $N$ with $\lambda_{PMS}^{(1)}$ (second column) and 
with $\lambda=0$ (third column). The underlined digits are correct.}
\begin{ruledtabular}
\begin{tabular}{ccc}
$N$&$\zeta_{PMS}(2,3/2,1)$&$\zeta_{\lambda=0}(2,3/2,1)$\\
\hline
1   & \underline{1}.4728636067646540152245808470808148333385501418201 & 0.64666527044453939590268993182589873867936491308763\\
10  & \underline{1.5124}141548516402130603869952574426525740627123155 & 3.4054827271508252417934796716746660253808017949304\\
100 & \underline{1.5124349215502030648030954892350003}557973564160761 & 7.189509053068942853725754726982871981048977697327\\
200 & \underline{1.5124349215502030648030954892350003618366746247481} & 9.5161848472956355375285584079161971312864869056589 \\
\end{tabular}
\end{ruledtabular}
\end{table*}

As an example of the application of our formula, we consider the case studied in eq.~(12) of \cite{LV97}, i.e.
\beq
I(m) &=& T \ \mu^{2 \epsilon} \ \sum_{n=-\infty}^{+\infty} \int \frac{d^Dk}{(2\pi)^D}  \ \frac{1}{k^2+M^2} \nonumber \\
&=& T \ \mu^{2 \epsilon} \  \int \frac{d^Dk}{(2\pi)^D}  \ \frac{1}{k^2+M^2} + J(M,0,1) 
\eeq
which is essentially the one-loop self energy.

The first integral is finite and equal to $- M T/(4 \pi)$ whereas the second term can be written as:
\begin{widetext}
\beq
J(M,0,1) 
 &=&  2^{-2-2\epsilon} \ \pi^{-3/2- 3 \epsilon} \ \mu^{2\epsilon} \ T^{1-4\epsilon} \ 
\Gamma(-1/2+\epsilon) \ \left[  - \frac{1}{M^{-1-2\epsilon}} + \zeta(2,-1/2+\epsilon,M^2)\right]  \ ,
\eeq
\end{widetext}
where the divergencies are contained in $\zeta(2,-1/2+\epsilon,M^2)$.

By using our eq.~(\ref{l0}) and retaining only the divergent terms and those independent of $\epsilon$ we 
obtain
\beq
I(m) &=& - \frac{m T}{4 \pi} + \frac{T^2}{12} - \frac{m^2}{16 \pi^2} \ \left[\frac{1}{\epsilon}  + \gamma - 
\log \frac{4\pi T^2}{\mu^2} \right] \nonumber \\
&+& \frac{m^4 \zeta(3)}{8 (2\pi)^4 T^2} 
-   \frac{m^6\,\zeta(5)}{1024\,{\pi }^6\,T^4} + 
  \frac{5\,m^8\,\zeta(7)}{32768\,{\pi }^8\,T^6}  \nonumber \\
&-&   \frac{7\,m^{10}\,\zeta(9)}{262144\,{\pi }^{10}\,T^8} + \dots
\eeq
which reproduces the results of \cite{LV97}, which, however, were considered 
only up to order $m^4$. As anticipated, the formula obtained is a polynomial in $m$.
 
We will now use the optimal series (\ref{eqq}) to improve this result. We first notice that 
since
\beq
\frac{d \zeta(u,s,\xi)}{d\xi}  &=& - s \ \zeta(u,s+1,\xi)
\label{eqder}
\eeq
a divergent series can be related to a convergent series by taking repeated derivatives 
with respect to the parameter $\xi$. Indeed by applying  eq.~(\ref{eqder}) twice we can write
the general expression 
\begin{widetext}
\beq
J(M,a,b) &=& \frac{2^{1 + 2\,a - 2\,b + 2\,{\epsilon}}\,
    {{\Omega}}^{2\,{\epsilon}}\,
    {\pi }^{\frac{3}{2} + 2\,a - 2\,b + {\epsilon}}\,
    T^{4 + 2\,a - 2\,b}\,\Gamma(\frac{3}{2} + a - 
      {\epsilon})\,\Gamma(-\left( \frac{3}{2} \right)  - 
      a + b + {\epsilon})}{\Gamma(b)\,
    \Gamma(\frac{3}{2} - {\epsilon})} \nonumber\\
&\cdot& \left( 1 + a - b + \frac{3 - 2\,{\epsilon}}{2} \right) \,
  \left( -a + b + \frac{-3 + 2\,{\epsilon}}{2} \right) \nonumber \\
&\cdot&\left\{ \int dM^2 \int dM^2
  \left( -M^{-2\,\left( 2 - a + b + \frac{-3 + 2\,{\epsilon}}{2} \
\right) } + \zeta(2,\frac{1}{2} - a + b + \epsilon,M^2) \right)  \right\}
\eeq
\end{widetext}
where the last $\zeta$ function is now fully convergent when $a=0$ and $b=1$. Since this formula requires a double
integration of a convergent series, the result is determined up to two constants of integration, which will
contain the divergent contributions.

By applying eq.~(\ref{eqq}) to first order and using the optimal value of $\lambda$ given by eq.~(\ref{s3_7}) we
obtain the  simple result
\begin{widetext}\beq
J_{PMS}(M,0,1) &=&   T^2 \ \left( - \kappa_1 \ M^2   - \frac{{\zeta(3)}^{5/2} \
       {\sqrt{\zeta(3) + M^2\, \zeta(5)}}}{{\zeta(5)}^2} \right) + \kappa_2 \nonumber \\
\label{jpms0}
\eeq\end{widetext}
where $\kappa_{1,2}$ are the constants of integration {\sl independent} of $M$.
In order to determine these constants we Taylor expand this expression in $M$ and then use
$M = m/(2 \pi T)$ thus obtaining 
\beq
J(M,0,1) &\approx&  \kappa_2 - \frac{\left( \kappa_1 \,m^2 \right) }{4\,{\pi }^2} + 
  \frac{m^4\,\zeta(3)}{128\,{\pi }^4\,T^2} - 
  \frac{T^2\,{\zeta(3)}^3}{{\zeta(5)}^2} \nonumber \\
&-& \frac{m^2\,{\zeta(3)}^2}{8\,{\pi }^2\,\zeta(5)} -   
\frac{m^6\,\zeta(5)}{1024\,{\pi }^6\,T^4} + 
  \frac{5\,m^8\,{\zeta(5)}^2}
   {32768\,{\pi }^8\,T^6\,\zeta(3)} \nonumber \\
&-& 
  \frac{7\,m^{10}\,{\zeta(5)}^3}
   {262144\,{\pi }^{10}\,T^8\,{\zeta(3)}^2} + \dots
\eeq

This expression can be now compared with the perturbative result and the constants of integration 
can thus be extracted
\beq
\kappa_1 &=& \frac{1}{4} \left[ \frac{1}{\epsilon} + \gamma - \log \left(\frac{4 \pi T^2}{\mu^2}\right)-
\frac{2\,{{\zeta}(3)}^2}{{\zeta}(5)}\right] \nonumber \\
\kappa_2 &=& \frac{1}{12} + \frac{{{\zeta}(3)}^3}{{{\zeta}(5)}^2} \nonumber \ .
\eeq

Eq.~(\ref{jpms0}) is a quite remarkable formula: indeed, although it has been derived by applying our 
method to first order, it reproduces correctly  the terms going as $m^4$ and $m^6$ and it also provides the 
coefficients of the higher order terms apart from factors depending on the Riemann $\zeta$ function evaluated 
at odd integer values~\footnote{We have checked this property to much higher order than the ones reached in the formula.}. 
Of course, we can let the perturbative result guide us further and use the fact that $\lim\limits_{n\rightarrow \infty} 
\zeta(n)=1$ to write the improved formula
\begin{widetext}
\beq
J_{PMS2}(M,0,1) &=& \frac{13}{12} \ T^2 - \frac{M^2\,T^2}{4} \ \left[ \frac{1}{\epsilon} + \gamma - \log \left(\frac{4 \pi T^2}{\mu^2}\right)-2
\right] \nonumber \\
&-& \frac{T^2}{16} \ \left[ \left( 16\,{\sqrt{1 + M^2}} - 2\,M^4\,\left( -1 + \zeta(3) \right)  + M^6\,\left( -1 + \zeta(5) \right)  \right)  \right] \ ,
\label{jpms2}
\eeq
\end{widetext}
where the notation $PMS2$ has been introduced to distiguish it from the previous formula.
Notice that both equations (\ref{jpms0}) and (\ref{jpms2}) are non--polynomial in $m$.

By putting the pieces together we finally obtain our approximation to $I(m)$, given by
\begin{widetext}\beq
I_{PMS2}(M) &=& - \frac{M T^2}{2 } +
\frac{13}{12} \ T^2 - \frac{M^2\,T^2}{4} \ \left[ \frac{1}{\epsilon} + \gamma - \log \left(\frac{4 \pi T^2}{\mu^2}\right)-2
\right] \nonumber \\
&-& \frac{T^2}{16} \ \left[ \left( 16\,{\sqrt{1 + M^2}} - 2\,M^4\,\left( -1 + \zeta(3) \right)  + M^6\,\left( -1 + \zeta(5) \right)  \right)  \right] \ .
\label{ipms2}
\eeq
\end{widetext}
When negative values of $m^2$ are considered, corresponding to a spontaneously broken phase, 
these results hold for $T \geq T_0 \equiv - m^2/(4 \pi^2)$.

As a further application of the method described in this paper we  consider the integral
\beq
{\cal J}(\phi) = \frac{T}{2} \mu^{-2\epsilon}  \sum_{k_0} \int \frac{d^{3-2\epsilon}k}{(2\pi)^{3-2\epsilon}} \
\log \left( k^2+m^2(\phi)\right)
\eeq
where $k_0 = 2 \pi n T$, for all $n$ integers. The effective potential to one--loop is expressed in terms of this
integral.

Given the relation
\beq
I(m) &=& \frac{1}{m} \  \frac{d{\cal J}}{dm}
\eeq
it is not necessary to calculate  ${\cal J}(m)$ which can be obtained as
\beq
{\cal J}(m) = \int I(m) \ m \ dm + \rho  \ ,
\eeq
where $\rho$ is constant of integration independent of $m$.

By using $I_{PMS2}$ of eq.~(\ref{ipms2}) we obtain the simple expression:
\beq
{\cal J}_{PMS2}(m) &=& \rho + \frac{13}{24} \,m^2\,T^2 -  \frac{m^3\,T}{12\,\pi } \nonumber \\
&-& \frac{T}{6 \pi} \ \left( m^2+4\ \pi^2 T^2\right)^{3/2} \nonumber \\
&-& \frac{m^4}{64\pi^2} \ \left[ \frac{1}{\epsilon} + \gamma - \log \left(\frac{4 \pi T^2}{\mu^2}\right)-2 \right] \nonumber \\
&+& \frac{m^6}{768 \ \pi^4 T^2} \ \left(\zeta(3)-1\right) \nonumber \\
&-& \frac{m^8}{8192 \ \pi^6 T^4} \ \left(\zeta(5)-1\right)
\label{jpms}
\eeq
which can be expanded in powers of $m$ to give
\beq
{\cal J}_{PMS2}(m) &=&  \rho - \frac{4\,{\pi }^2\,T^4}{3} + \frac{m^2\,T^2}{24}- \frac{m^3\,T}{12\,\pi }  \nonumber \\
&-& \frac{m^4}{64 \pi^2} \,\left( \frac{1}{\epsilon} + \gamma - \log \left(\frac{4 \pi T^2}{\mu^2}\right) \right)\nonumber \\
&+& \frac{m^6 \zeta(3)}{768\,{\pi }^4\,T^2}  - \frac{m^8 \zeta(5)}{8192\,{\pi }^6\,T^4} \nonumber \\
  &+& \frac{m^{10}}{65536\,{\pi }^8\,T^6} + \dots
\eeq
which reproduces exactly eq.~(3.7) of \cite{AE93}, provided that $\rho=\frac{4 \pi^2 T^4}{3} + constant$. This last constant is 
independent both of $m$ and of $T$.

We are now in a position to compare our result of eq.~(\ref{jpms}) with the high temperature expansion of eq.(42) of ref.~\cite{HW82}
which reads\footnote{We divide by an overall factor of $2$.}
\beq
\frac{\Omega}{2 V T^4} &=& - \frac{\pi^2}{90} + \frac{y^2}{6} - \frac{y^3}{12 \pi} + \frac{y^4}{32 \pi^2} \ 
\left[ \log \frac{4\pi}{y} - \gamma + \frac{3}{4} \right] \nonumber \\
 &-& \frac{y^4}{16 \pi^2}  \sum_{k=1}^\infty 
(-1)^k \ \left(\frac{y}{4 \pi}\right)^k \ \frac{\Gamma(2 k +1) \ \zeta(2 k+1)}{\Gamma(k+1) \ \Gamma(k+3)} \nonumber \\
\label{hw}
\eeq
after setting the chemical potential to zero\footnote{In such limit the hypergeometric functions in the original formula
of \cite{HW82} all go to $1$.}. We have defined $y = m/T$. 

As we can see from table~\ref{tab:table2} the result of \cite{LV97} for $I(m)$ would essentially provide the expansion 
of eq.~(\ref{hw}) and thus reproduce the results obtained long time ago by Haber and Weldon~\cite{HW82}. 
We regard this procedure as ``perturbative'' meaning that, when the series
in eq.~(\ref{hw}) is truncated a polynomial in powers of $m/T$ is obtained. On the other hand, our simple formula, 
eq.~(\ref{jpms}) reproduces correctly the perturbative expansion of \cite{HW82} up to order $(m/T)^8$, and up to a factor 
involving the $\zeta$ function calculated at odd integer values, which however tend to $1$  for large values. 
This makes our simple formula quite precise. 
We remind the reader that eq.~(\ref{jpms}) was obtained by applying our method to {\sl first order} and that, given the
convergence of the method, drastic improvements in the quality of the approximation are expected if higher orders would to
be taken into account. Although the calculation of higher orders with our method  would be an interesting issue by itself, we will leave it for
 future work.

\begin{figure}
\includegraphics[width=8cm]{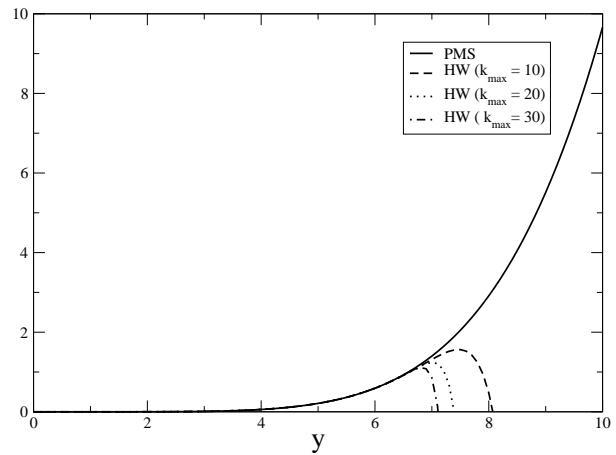}
\caption{Comparison between  $\tilde{{\cal J}}_{PMS2}/T^4$ and eq.~(\ref{hw}) taking $k_{max} = 10$ (dashed), $k_{max} = 20$  (dotted)
and $k_{max} = 30$  (dot-dashed) terms.
 \label{Fig_2}}
\end{figure}

In fig.~\ref{Fig_2} we have plotted $\tilde{{\cal J}}_{PMS2}/T^4$, defined by taking out the terms up to order
$(m/T)^4$, and the similar functions obtained from eq.~(\ref{hw}),  performing the sum to three different orders ($k_{max} = 10,20,30$).
The horizontal scale is $y = m/T$. We leave to the reader any judgement on the quality of our approximation.
Notice that the series of Haber and Weldon should correspond to our variational series for $\lambda = 0$; as we have seen before
the criterion of convergence for a series is that $\lambda^2 > \frac{M^2-1}{2}$, which for $\lambda=0$ can be fullfilled only if
$m/T < 2 \pi$.  Looking at the figure we indeed see that when we get close to this value the sum of \cite{HW82} becomes ill--behaved.

\begin{table}
\caption{\label{tab:table2} Comparison between eq.~(\ref{hw}) and the simple approximation of eq.~(\ref{jpms}).}

\begin{ruledtabular}
\begin{tabular}{ccc}
$N$& eq.~(\ref{hw}) & ${\cal J}_{PMS2}/T^4$\\
\hline
$\left[\frac{m}{T}\right]^6$   & $\frac{\zeta(3)}{768 \ \pi^4}$ & $\frac{\zeta(3)}{768\, \pi ^4}$ \\
$\left[\frac{m}{T}\right]^8$  & $-\frac{\zeta(5)}{8192 \ \pi^6}$ & $-\frac{\zeta(5)}{8192 \ \pi^6}$ \\
$\left[\frac{m}{T}\right]^{10}$ & $\frac{\zeta(7)}{65536 \ \pi^8}$ &  $\frac{1}{65536 \ \pi^8}$ \\
$\left[\frac{m}{T}\right]^{12}$  & $-\frac{7 \ \zeta(9)}{3145728 \ \pi^{10}}$ &   $-\frac{7}{3145728 \ \pi^{10}}$ \\
$\left[\frac{m}{T}\right]^{14}$  & $\frac{3 \ \zeta(11)}{8388608 \ \pi^{12}}$ &  $\frac{3}{8388608 \ \pi^{12}}$  \\
\end{tabular}
\end{ruledtabular}
\end{table}

\section{Conclusions}
\label{concl}

The method that we have described in this paper is quite general and probably could be extended in the future to deal with 
a larger class of problems than the one presented here. We have proved that, by using variational techniques it is possible to
estimate analytically and to an arbitrary degree of precision series which are difficult to evaluate with standard techniques.
In particular, we have shown that by relating divergent series to convergent ones through repeated derivatives and then matching 
the divergences contained in the constants of integration with the ones coming out of the perturbative calculation, it is possible 
to construct a really non--perturbative regularization. Such regularized expressions are very accurate even at low orders and low temperatures.
There is a huge literature dealing with field theoretical problems at finite temperature and we feel that this paper can provide
a quite general and useful tool to attack many of these problems. We also believe, although at present is still to be confirmed, that 
the method that we have described could be useful in the non--perturbative calculation of the Casimir effect, for which zeta function
regularization is a well--established technique (see for example \cite{Ez02}). It would be quite interesting to see if non--perturbative
expressions for the  Casimir effect could be calculated analytically. We hope to apply these ideas to such a problem in the near future.
As a final remark, we like to stress that despite the simplicity of the ideas that we have illustrated, all the results that we have
obtained are {\sl fully analytical} and improvable to the desired level of accuracy.


\begin{acknowledgments}
The author acknowledges support of Conacyt grant no. C01-40633/A-1 and of the Fondo Ram\'on Alvarez Buylla of 
Colima University. I thank Dr. Alfredo Aranda and Dr. Hugh Jones for reading this manuscript. In particular I am indebted with Prof. Jones
for pointing out several typos and for suggesting the simpler form of eq.~(\ref{ap_1_8}).
\end{acknowledgments}

\end{document}